\documentclass[
superscriptaddress,
groupedaddress,
twocolumn,
 amsmath,amssymb,
 aps,
prx,
]{revtex4}

\usepackage{graphicx}
\usepackage{dcolumn}
\usepackage{bm}
\usepackage{amsmath, esint}
\usepackage{amssymb}
\usepackage{physics}
\usepackage{bigints}

\DeclareMathAlphabet\mathbfcal{OMS}{cmsy}{b}{n}

\newcommand{\Id}{\mathbb{I}}

\begin{document}

\preprint{APS/123-QED}

\newcommand\numthis{\stepcounter{equation}\tag{\theequation}}
\newcommand{\im}[1]{\operatorname{Im}\left[#1\right]}
\newcommand{\re}[1]{\operatorname{Re}\left[#1\right]}

\title{Fundamental limits on $\chi^{(2)}$ second harmonic generation}
\date{July 2023}

\author{Jewel Mohajan}
\affiliation{Department of Electrical and Computer Engineering, Princeton University, Princeton, New Jersey 08544, USA}

\author{Pengning Chao}
\affiliation{Department of Electrical and Computer Engineering, Princeton University, Princeton, New Jersey 08544, USA}

\author{Weiliang Jin}
\affiliation{Flexcompute Inc., 130 Trapelo Road, Belmont, Massachusetts 02478, USA}

\author{Sean Molesky}
\affiliation{Department of Engineering Physics, Polytechnique Montréal, Montréal, Québec H3T 1J4, Canada}

\author{Alejandro W. Rodriguez}
\affiliation{Department of Electrical and Computer Engineering, Princeton University, Princeton, New Jersey 08544, USA}

\begin{abstract}
Recent advances in fundamental performance limits for power quantities based on Lagrange duality are proving to be a powerful theoretical tool for understanding electromagnetic wave phenomena. To date, however, in any approach seeking to enforce a high degree of physical reality, the linearity of the wave equation plays a critical role. In this manuscript, we generalize the current quadratically constrained quadratic program framework for evaluating linear photonics limits to incorporate nonlinear processes under the undepleted pump approximation. Via the exemplary objective of enhancing second harmonic generation in a (free-form) wavelength-scale structure, we illustrate a model constraint scheme that can be used in conjunction with standard convex relaxations to bound performance in the presence of nonlinear dynamics. Representative bounds are found to anticipate features observed in optimized structures discovered via computational inverse design. The formulation can be straightforwardly modified to treat other frequency-conversion processes, including Raman scattering and four-wave mixing.
\end{abstract}

\maketitle

Structural optimization has emerged as a new paradigm for the design of physical devices subject to wave and scattering physics. The typical electromagnetic wave problem consists of optimizing some quadratic field objective (e.g., the energy density at some location or the power radiated by a source) with respect to structural variations of the scattering medium. Because such design problems are non-convex, there are generally no guarantees of convergence to optimal solutions. Recently, however, a new means of probing wave equations makes possible the calculation of physical limits, or performance bounds, transforming the non-convex structural design problem into a convex and structure-agnostic field optimization problem with controlled relaxations of wave physics. Relying on the linearity of the wave equation, a bound on any quadratic wave objective is formulated as the solution of a quadratically constrained quadratic program (QCQP) by means of convex relaxations, e.g., Lagrangian duality or semi-definite programming~\cite{chao_physical_2022,angeris_heuristic_2021,gertler_many_2023,gustafsson_upper_2020}. This general framework has enjoyed great success in establishing new limits and scaling laws for diverse electromagnetic design objectives, from scattering cross sections to local density of states and focusing. Thus far, however, no such approach for deriving fundamental limits in the context of nonlinear photonics has been developed. Establishing limits beyond linear physics is of importance not only in guiding ongoing investigations of compact lasers, quantum controllers, and power limiters, but also in other disciplines including microfluidics and quantum mechanics. 

In this paper, we propose a Lagrange duality framework~\cite{chao_physical_2022} that is applicable to broad classes of nonlinear photonics objectives. Focusing on the problem of light incident on a $\chi^{(2)}$ or Pockels medium, we show that under appropriate modifications, the introduction of auxiliary degrees of freedom transforms this nonlinear wave optimization problem into a QCQP susceptible to bound calculation through Lagrangian dual relaxations. Technically, proper selection of convex physical constraints is needed to ensure a feasible dual problem and non-trivial bounds. We use this approach to investigate upper limits to second harmonic generation (SHG), the maximum power that may be generated at twice the oscillating frequency of a harmonic source interacting with a Pockels medium. Apart from establishing limits for second harmonic power objectives, the bounds can predict the scalings and trends one may expect from optimal structures with respect to changes in material and device size. The method allows systematically increasing the number of physical constraints incorporated to achieve tighter limits.  Finally, we propose several fabrication-ready doubly resonant slab cavities exhibiting among the highest nonlinear mode overlaps found in the literature. Extrapolating the observed performance gap in proof-of-concept two-dimensional settings to these realistic structures suggests little room for additional improvements beyond what is seen in topology-optimized cavities.

Second harmonic generation is an important phenomenon in nonlinear optics, with many practical applications including high frequency source generation~\cite{chen_design_1995,chen_high-performance_2021}, electro-optic modulators~\cite{liu_review_2015}, as well as high resolution imaging~\cite{pavone_second_2014} and spectroscopy~\cite{heinz_spectroscopy_1982,wang_second_2019}, to name a few. While bulk nonlinearities are weak, the output power from a field incident on a nonlinear medium at the second harmonic can be resonantly enhanced through structuring: a cavity supporting tightly confined modes at both wavelengths leads to higher field intensities, or smaller cavity mode volumes $V=\frac{\int \varepsilon \abs{\vb{E}}^2 \ d\vb{r}}{\max{\left(\varepsilon \abs{\vb{E}}^2\right)}}$, and longer interaction times, or larger quality factors $Q$. The resulting enhancement is captured by a figure of merit for the generated power at the second harmonic, $\eta \propto |\beta|^2 Q_{1}^2 Q_2$, proportional to the temporal lifetimes of both fundamental $Q_1$ and second harmonic $Q_2$ resonances, and a nonlinear field-overlap factor $\beta = \frac{1}{4}\frac{\int \varepsilon_0\sum_{ijk}\chi^{(2)}_{ijk} \left( {E}_{1i}^{*}{E}_{1j}^{*}{E}_{2k}+ {E}_{1i}^{*}{E}_{2j}{E}_{1k}^{*}\right) \ d\vb{r} }{\left( \int \varepsilon_0\varepsilon_1|\mathbf{E}_1|^2 \ d\vb{r}\right) \sqrt{\int \varepsilon_0\varepsilon_2|\mathbf{E}_2|^2\ d\vb{r}} }$, where $\varepsilon_1$ and $\varepsilon_2$ denote the relative permittivity of the structure at the fundamental and second harmonic wavelengths, while $\vb{E}_1$ and $\vb{E}_2$ represent the corresponding mode profiles~\cite{lin_cavity-enhanced_2016,rodriguez_2_2007}; $\beta$ scales $\propto 1/\sqrt{V}$ and c taptures the degree of spatial confinement and mode coupling. The problem of designing resonators maximizing $\eta$ is complicated because $Q$ and $\beta$ are often interconnected, with smaller cavities leading to larger radiative losses~\cite{chao_maximum_2022}. (Note that photonic crystals often exhibit a single narrow gap, precluding use of gap confinement at multiple wavelengths~\cite{joannopoulos_photonic_2008}.) Previously, the relative ease in conceptualizing and fabricating large-etalon resonators supporting a plethora of tunable modes, proposed in the mid 1960s~\cite{fejer_nonlinear_1994}, led to devices with slower operation (larger $Q$) and smaller $\beta$. Over the past few decades, however, fabrication advances and the growing need to miniaturize devices (on a microchip) along with the promise of larger nonlinearities and greater operating speeds ushered a search for compact microcavities, from millimeter-\cite{furst_naturally_2010} and micrometer-scale whispering gallery mode resonators ~\cite{bi_high-efficiency_2012,logan_400w_2018,pernice_second_2012} to, more recently, wavelength-scale structures designed via brute-force optimization~\cite{lin_cavity-enhanced_2016}. In going toward smaller devices, the chief challenge of achieving simultaneous resonances at far-apart wavelengths poses a conceptual limitation~\cite{bravo-abad_efficient_2010,rodriguez_2_2007}, making quantitative analyses of expected trade-offs in temporal and spatial confinement very challenging. Recent inverse designs like the novel class of slab geometry proposed below, appear to exploit a range of physical mechanisms---from index guiding to Bragg scattering and slot confinement---to create spectrally far-apart, highly spatially confined resonances with moderately large bandwidths. This paper represents a top-down engineering perspective on the problem of enhancing frequency conversion in structured environments, setting fundamental wave constraints that complement our recent bottom-up investigations based on large-scale structural optimization~\cite{lin_cavity-enhanced_2016}.

We note that while there have been prior attempts at establishing bounds on Raman scattering of single molecules~\cite{michon_limits_2019}, those bounds exploit special simplifying characteristics of that problem---the nonlinear process occurs at a single fixed point in space---to factor the nonlinear photonics objective into a product of two linear objectives; such a factorization is not possible in more general settings such as problems involving extended nonlinear media. Independent linear photonics bounds were then evaluated using solely the constraint of passivity~\cite{miller_fundamental_2016} and multiplied together to obtain a bound on the Raman scattering power; this neglects potential trade-offs in performance when simultaneously optimizing both linear objectives~\cite{molesky_t_2022}, and passivity alone does not adequately capture multiple scattering effects in photonic structures~\cite{molesky_global_2020,chao_maximum_2022}. The framework presented below allows for a more complete account of wave physics as embodied by Maxwell's equations~\cite{chao_physical_2022} and consideration of spatially distributed nonlinearities subject to the typical restrictions of finite resonance strength and non-depletion of the pump; hence, it is generalizable to many other wave objectives and nonlinear processes.
\begin{figure}[htp]
\centering
\includegraphics[width=\linewidth]{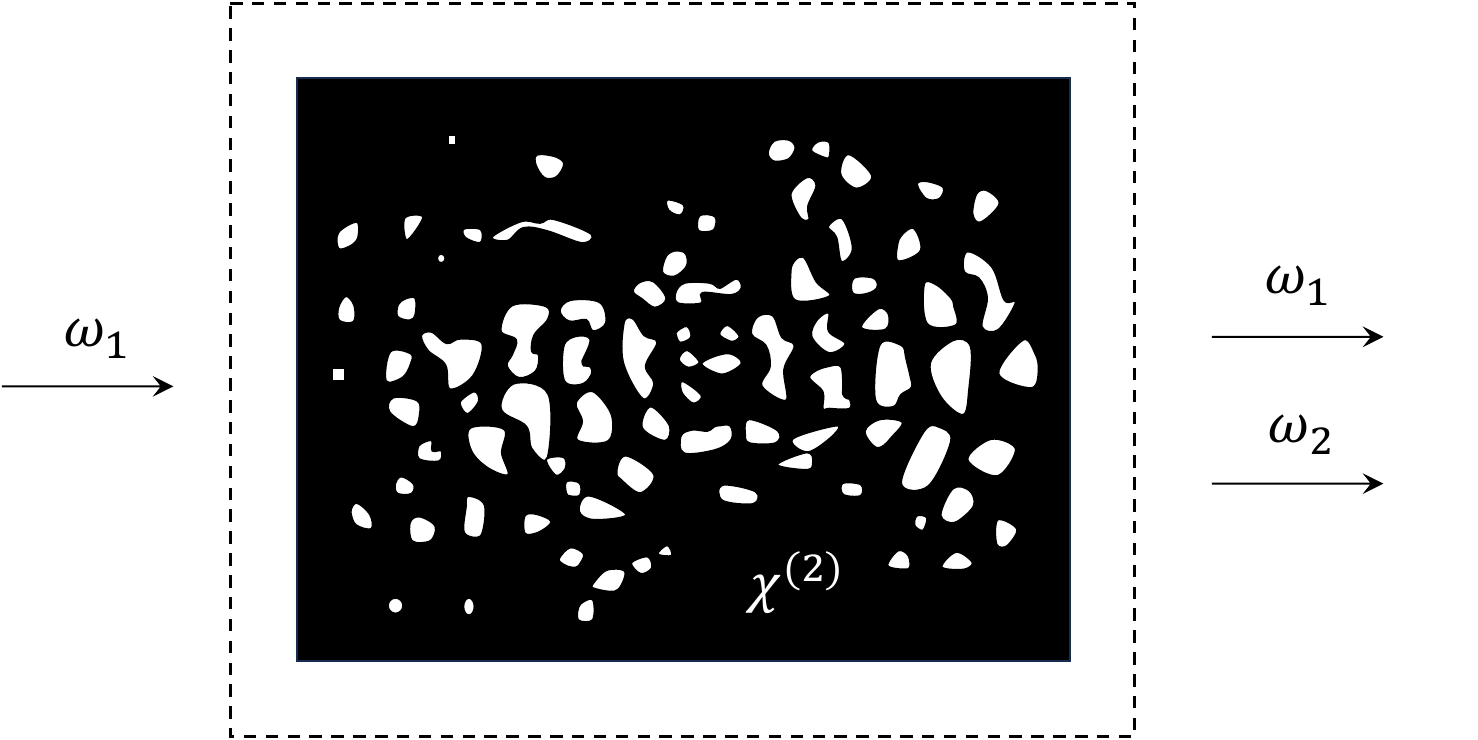}
\caption{Schematic illustration of the second-harmonic generation problem under investigation: a device (representative design obtained through structural optimization) contained within some design region efficiently converts light incident at $\omega_1$ to the second harmonic frequency $\omega_2=2\omega_1$ by addressing the multi-prong challenge of supporting two spectrally far-apart and long-lived but tightly confined and spatially overlapping modes~\cite{lin_cavity-enhanced_2016}. The proposed optimization framework establishes shape-independent limits on the largest achievable conversion efficiencies that may be realized for a given material and design region.}
\label{fig:nl}
\end{figure}

\textbf{Formulation---}Consider a structure represented by spatially varying linear susceptibilities $\overline{\chi}_j(\vb{r}) = \chi_j \overline{\rho}(\vb{r})$ for $j=\{1,2\}$, at both fundamental $\omega_1$ and second-harmonic $\omega_2 = 2\omega_1$ frequencies, and by a nonlinear susceptibility $\overline{\chi}^{(2)}(\vb{r}) = \chi^{(2)} \overline{\rho}(\vb{r})$. Here, $\overline{\rho}(\vb{r})$ denotes the distribution of material and is a scalar field that is allowed to vary continuously from $0$ to $1$ within some design volume $\Omega$, while $\chi_1$, $\chi_2$, and $\chi^{(2)}$ are tensors representing the bulk susceptibilities of the medium comprising the structure. The design domain is illuminated by an incident electric field $\vb{E}_1^{inc}(\vb{r})$ created by a current source $\vb{J}^{inc}(\vb{r})$ oscillating at frequency $\omega_1$. For brevity, we refer to $\chi$ as the bulk material susceptibility and $\overline{\chi}$ as its spatial distribution, and omit spatial dependencies of both fields and distributions for the remainder of the manuscript. Generally, this nonlinear problem is challenging to solve, leading for instance to an infinite sequence of ``up-converted'' and ``down-converted'' waves at multiples of the incident frequency. However, the weak nature of nonlinearities allows for several simplifications. First, the scarcity and challenge of designing perfectly matched resonances at several, far-apart wavelengths means that one can focus exclusively on energy transfer between the fundamental ($\vb{E}_1$) and second harmonic ($\vb{E}_2$) fields. Second, under the undepleted pump approximation~\cite{boyd_nonlinear_2008}, one may neglect the down-converted field at the fundamental frequency. Finally, to lowest order in the nonlinearity, one may treat the induced nonlinear polarization at the second harmonic as a free source generating fields at $\omega_2$. These simplifying assumptions made it recently possible to frame the problem of maximizing $\eta$ as a sequence of coupled scattering problems~\cite{lin_cavity-enhanced_2016}, described by the following structural optimization program:
\begin{equation}\label{eq:opt1}
\begin{aligned}
\max_{\overline{\rho}} \quad &  -\frac{1}{2} \re{\int_\Omega \vb{J}_2^* \cdot \vb{E}_2 \ d\vb{r}}\\
\textrm{s.t.} \quad & \mathbb{M}_{1} \vb{E}_1 = \overline{\chi}_{1}\vb{E}_1 +  \frac{i}{\omega_1 \varepsilon_0} \vb{J}_1 + \frac{i}{\omega_1\varepsilon_0} \vb{J}^{inc}\\
&\mathbb{M}_{2} \vb{E}_2 = \overline{\chi}_{2} \vb{E}_2 + \frac{i}{\omega_2 \varepsilon_0}\vb{J}_2\\
&\vb{J}_1 = -i\omega_1\varepsilon_0\overline{\chi}^{(2)} (\vb{E}_1^* \odot \vb{E}_2) \\ 
&\vb{J}_2 = -i\omega_2\varepsilon_0\overline{\chi}^{(2)}(\vb{E}_1 \odot \vb{E}_1)\\
&\overline{\chi} = \chi \overline{\rho}, \qquad \overline{\rho} \in [0,1].  
\end{aligned}
\end{equation}
Here, the superscript $( ^*)$ denotes conjugated vector fields, and ($\odot$) denotes the Hadamard product (element-wise multiplication). $\varepsilon_0$ is the vacuum permittivity and $\mu_0$ the permeability of the free space. For simplicity, we assume that all linear and nonlinear material susceptibilities are local and isotropic, i.e., they can be expressed as simple scalars. The $e^{-i\omega t}$ time convention is used to represent harmonic fields, with the vacuum Maxwell operator given by $\mathbb{M}_{l} = \left(-\mathbb{I}+ \frac{1}{\omega_l^2 \varepsilon_0}\curl\frac{1}{\mu_0}\curl\right), \ l={1,2}$; where $\mathbb{I}$ is the identity operator. Finally, we defined $\vb{J}_1$ and $\vb{J}_2$ as the induced currents resulting from nonlinear down-conversion and up-conversion, respectively. 

In line with the arguments presented in Ref.~\cite{lin_cavity-enhanced_2016}, $\vb{J}_2$ may be considered here as a free source at $\omega_2$ while $\vb{J}_1$, included in (\ref{eq:opt1}) for technical reasons discussed further below, may be omitted under the undepleted pump approximation. Ignoring down-conversion, (\ref{eq:opt1}) amounts to the maximization of a quadratic objective satisfying a set of coupled partial differential equations, the solution of which maximizes $\eta \propto |\beta|^2 Q_{1}^2 Q_2$. Effectively, by Poynting's theorem, $-\frac{1}{2} \re{\int_\Omega \vb{J}_2^* \cdot \vb{E}_2 \ d\vb{r}}$ is the power extracted from the polarization field at the second harmonic. Writing $\mathbf{J}_2$ explicitly in terms of the fundamental field in the objective yields the desired proportionality to $\beta$. Moreover, supposing structures and fields consistent with resonant enhancement, maximizing extracted power at second harmonic yields the desired scaling with respect to $Q_1$ and $Q_2$~\cite{lin_cavity-enhanced_2016}. Note that in principle, one may also consider other power quantities as metrics of enhanced SHG, such as the net scattered/radiated power at the second harmonic given by $\Phi_2 \equiv -\frac{1}{2} \re{\int_\Omega \vb{J}_2^* \cdot \vb{E}_2 \ d\vb{r}} - \frac{1}{2}\varepsilon_0 \omega_2 \im{\int_\Omega \vb{E}_2^* \cdot \overline{\chi}_2 \vb{E}_2 \ d\vb{r}}$, which we exploit further below. Even under these simplifications, several difficulties must be addressed before one can arrive at a corresponding bound optimization problem, chief among them being the need to derive an analogous QCQP with dual feasibility. As detailed in \cite{chao_physical_2022} and further below, the transformation from a structural to a bound optimization problem requires shifting from structural $\overline{\rho}$ to polarization degrees of freedom, which leads to an objective that is evidently not quadratic. We solve this problem by promoting $\vb{J}_2$ from a free source to an optimization degree of freedom, enforcing the nonlinear relation $\vb{J}_2 = -i\omega_2\varepsilon_0 \overline{\chi}^{(2)}(\vb{E}_1 \odot \vb{E}_1)$ instead as an auxiliary constraint. Additional relaxations and modifications are needed to ensure that this auxiliary constraint leads to tractable dual solutions. Finally, we show that including the down-conversion term in (\ref{eq:opt1}) leads to a power conservation constraint that is convex for all polarization fields, which makes finding feasible points for the dual straightforward. Further details follow below.

We begin by defining the following optimization degrees of freedom:
\begin{align}
    \vb{P}_1 &= \overline{\chi}_{1}\vb{E}_1\\
    \vb{P}^{(2)} &= \overline{\chi}^{(2)} \left(\vb{E}_1 \odot \vb{E}_1\right) \\
    \vb{P}_2 &= \overline{\chi}_{2} \vb{E}_2 + \vb{P}^{(2)}
\end{align}
with $\vb{P}_1$ denoting the induced polarization at $\omega_1$, and $\vb{P}^{(2)}$ and $\vb{P}_2$ the nonlinear and net polarization at $\omega_2$, respectively. Following a similar procedure as in \cite{molesky_hierarchical_2020,kuang_computational_2020}, instead of enforcing that any optimal polarization solves Maxwell’s equations everywhere, the problem is relaxed by imposing instead spatial integral relations over sub-domains of the design domain $\Omega_k \subseteq \Omega$: 
\begin{align}
    \int_{\Omega_k}\vb{P}_1^* \cdot \vb{E}_1^{inc} \ d\vb{r}= \int_{\Omega_k} \vb{P}_1^* \cdot \left( \chi_{1}^{-1} \mathbb{I} - \mathbb{G}_1 \right) \vb{P}_1 \ d\vb{r}, \label{eq:PoyntingFF}
    \\ 
    \int_{\Omega_k} \vb{P}_2^* \cdot \chi_{2}^{-1} \vb{P}^{(2)} \ d\vb{r}= \int_{\Omega_k} \vb{P}_2^* \cdot \left( \chi_{2}^{-1}\mathbb{I} - \mathbb{G}_2 \right) \vb{P}_2 \ d\vb{r}. \label{eq:PoyntingSH}
\end{align}
These volumetric identities are complex generalizations of the familiar Poynting’s theorem, establishing energy conservation across $\Omega_k$ and written here explicitly in terms of the bulk susceptibility ${\chi}_1$, and ${\chi}_2$, and vacuum electric Green’s function $\mathbb{G}_l = \mathbb{M}_l^{-1}$, $l=1,2$. $\Omega_k$ may be any sub-domain within the design domain, with the introduction of constraints over smaller sub-domains (down to the pixel level in a computational grid) forcing solutions to resolve wave physics at increasingly smaller length scales. 

Apart from (\ref{eq:PoyntingFF}) and (\ref{eq:PoyntingSH}) which enforce energy conservation at each frequency separately, one may also impose similar constraints across fundamental and second harmonic frequencies that reflect the fact that the shape of any structure is the same for both frequencies~\cite{molesky_t_2022,shim_fundamental_2021}:
\begin{align}
    \int_{\Omega_k}\vb{P}_2^* \cdot \vb{E}_1^{inc} \ d\vb{r} &= \int_{\Omega_k} \vb{P}_2^* \cdot \left( \chi_{1}^{-1} \mathbb{I} - \mathbb{G}_1 \right) \vb{P}_1 \ d\vb{r} \label{eq:Poynting21}
    \\
    \int_{\Omega_k}\vb{P}^{(2)*} \cdot \vb{E}_1^{inc} \ d\vb{r} &= \int_{\Omega_k} \vb{P}^{(2)*} \cdot \left( \chi_{1}^{-1} \mathbb{I} - \mathbb{G}_1 \right) \vb{P}_1 \ d\vb{r} \label{eq:Poynting31}
    \\
    \int_{\Omega_k} \vb{P}_1^* \cdot \chi_{2}^{-1} \vb{P}^{(2)} \ d\vb{r} &= \int_{\Omega_k} \vb{P}_1^* \cdot \left( \chi_{2}^{-1}\mathbb{I} - \mathbb{G}_2 \right) \vb{P}_2 \ d\vb{r} \label{eq:Poynting12}
\end{align}
Notably, both sets of constraints are quadratic functions of the fields. Further, as noted above, the objective of (\ref{eq:opt1}) is also manifestly quadratic if $\vb{P}^{(2)}$ is considered as an optimization degree of freedom and its nonlinear relation to the fields,
\begin{equation}\label{eq:nl1}
    \vb{P}^{(2)} = \overline{\chi}^{(2)}\left( \overline{\chi}_{1}^{-1} \vb{P}_1 \right) \odot \left( \overline{\chi}_{1}^{-1} \vb{P}_1 \right),
\end{equation}
enforced instead as an auxiliary constraint. While (\ref{eq:nl1}) may be imposed as a collection of quadratic constraints, directly imposing this relation turns out to be practically ineffective in the context of the Lagrangian dual formulation due to highly non-convex nature of the constraint, which involves a non-Hermitian unconjugated inner product, leading to a loss of strong duality and a negligible impact on the bound. To get around this issue, we propose an alternate convex relaxation of (\ref{eq:nl1}) based on bounding the norm of $\vb{P}^{(2)}$ over sub-regions $\Omega_j$:
\begin{align}\label{eq:nl2}
    \int_{\Omega_j} \vb{P}^{{(2)}*}\cdot \vb{P}^{(2)} \ d\vb{r} &= \abs{\frac{\chi^{(2)}}{\chi_{1}^2}}^2 \int_{\Omega_j} \vb{P}_1^{2*} \cdot \vb{P}_1^2 \ d\vb{r} \nonumber \\ &\leq \mathcal{R} 
    \abs{\frac{\chi^{(2)}}{\chi_{1}^2}}^2 \left(\int_{\Omega_j} \vb{P}_1^* \cdot \vb{P}_1 \ d\vb{r} \right)^2,
\end{align}
where the relaxation relies on the assumption of a discretized basis representation for the vector fields; consequently (\ref{eq:nl2}) includes an explicit resolution factor $\mathcal{R}$ that will generally depend on the discretization scheme used to carry out the integrals. $\int_{\Omega_j} \vb{P}_1^* \cdot \vb{P}_1 $ can now be further bounded by the following auxiliary optimization 
\begin{equation}\label{eq:aux_opt}
\begin{aligned}
\max_{\vb{P}_1} \quad &  \int_{\Omega_j} \vb{P}_1^* \cdot \vb{P}_1 \ d\vb{r}\\
\textrm{s.t.} \quad & \int_{\Omega_l} \vb{P}_1^* \cdot  \vb{E}_1^{inc} \ d\vb{r} = \int_{\Omega_l} \vb{P}_1^* \cdot \left( \chi_{1}^{-1}\mathbb{I} - \mathbb{G}_1 \right) \vb{P}_1 \ d\vb{r} \\
&\textrm{for a given collection } \{\Omega_l |  \Omega_l \subseteq \Omega\}.
\end{aligned}
\end{equation}
Finally, denoting $B_j$ as the Lagrangian dual bound on (\ref{eq:aux_opt}) yields the quadratic constraint 
\begin{equation}\label{eq:nlnorm}
    \int_{\Omega_j}\vb{P}^{(2)*} \cdot \vb{P}^{(2)} \ d\vb{r} \leq \mathcal{R} \abs{\frac{\chi^{(2)}}{\chi_{1}^2}}^2 B_j^2.
\end{equation} Note the complete freedom in choosing the set of sub-regions $\Omega_j$, $\Omega_k$, $\Omega_l$ 
associated with the different constraints. In the numerical results shown later, for any given (\ref{eq:aux_opt}), a large number of $\Omega_l$ sub-region constraints (up to the pixel level) are imposed to obtain the tightest $B_j$, which empirically contributes strongly to the tightness of the overall bounds. 

Putting together the aforementioned relaxations and modifications of the structural optimization problem (\ref{eq:opt1}), we arrive at the following transformed QCQP problem for maximizing SHG:
\begin{equation}\label{eq:opt2}
\begin{aligned}
\max_{\vb{P}_1,\vb{P}_2,\vb{P}^{(2)}} \Phi_2 &= \frac{1}{2} \omega_2 \varepsilon_0 \im{\int_\Omega \vb{P}_2^* \cdot \mathbb{G}_2 \vb{P}_2 \ d\vb{r}} \\
& \textrm{s.t.} \quad  \\ \int_{\Omega_k}\vb{P}_1^* \cdot \vb{E}_1^{inc} \ d\vb{r} &= \int_{\Omega_k} \vb{P}_1^* \cdot \left( \chi_{1}^{-1} \mathbb{I} - \mathbb{G}_1 \right) \vb{P}_1 \ d\vb{r}\\
\int_{\Omega_k} \vb{P}_2^* \cdot \chi_{2}^{-1} \vb{P}^{(2)} \ d\vb{r}&= \int_{\Omega_k} \vb{P}_2^* \cdot \left( \chi_{2}^{-1}\mathbb{I} - \mathbb{G}_2 \right) \vb{P}_2 \ d\vb{r}    \\
\int_{\Omega_k} \vb{P}_2^* \cdot \vb{E}_1^{inc} \ d\vb{r}&= \int_{\Omega_k} \vb{P}_2^* \cdot \left( \chi_{1}^{-1} \mathbb{I} - \mathbb{G}_1 \right) \vb{P}_1 \ d\vb{r}\\
\int_{\Omega_k}\vb{P}^{(2)*} \cdot \vb{E}_1^{inc} \ d\vb{r}&= \int_{\Omega_k} \vb{P}^{(2)*} \cdot \left( \chi_{1}^{-1} \mathbb{I} - \mathbb{G}_1 \right) \vb{P}_1 \ d\vb{r}\\
\int_{\Omega_k} \vb{P}_1^* \cdot \chi_{2}^{-1} \vb{P}^{(2)} \ d\vb{r}&= \int_{\Omega_k} \vb{P}_1^* \cdot \left( \chi_{2}^{-1}\mathbb{I} - \mathbb{G}_2 \right) \vb{P}_2 \ d\vb{r}\\
\int_{\Omega_j}\vb{P}^{(2)*} \cdot \vb{P}^{(2)} \ d\vb{r} &\leq \mathcal{R} \abs{\frac{\chi^{(2)}}{\chi_{1}^2}}^2 B_j^2.
\end{aligned}
\end{equation}
Technically, the objective in (\ref{eq:opt1}) is the net extracted power at the second harmonic, given by $\frac{1}{2} \omega_2 \varepsilon_0 \im{\int_\Omega \vb{P}^{(2)*} \cdot \mathbb{G}_2 \vb{P}_2 \ d\vb{r}}$ when expressed in terms of the polarization fields. However, for computational convenience, we consider the slightly modified objective $\Phi_2$ above that amounts to the radiated power or difference between extracted and absorbed powers at the second harmonic. Since (\ref{eq:opt2}) enforces fewer constraints than Maxwell's equations, any solution via convex relaxations, such as Lagrangian duality, constitutes a bound on the objective. 

The convex nature of the Lagrangian dual implies that so long as one can find a feasible point for the dual, application of gradient based algorithms guarantees convergence to the global optimum. To find an initial feasible point of the Lagrangian dual problem, it is convenient to have a constraint or a set of constraints in the primal problem that is convex for all primal optimization degrees of freedom, $\vb{P}_1$, $\vb{P}_2$, and $\vb{P}^{(2)}$. Assuming the nonlinear susceptibility $\chi^{(2)}$ to be real and to have full permutation symmetry~\cite{boyd_nonlinear_2008} and including the down-conversion term, albeit small, an elegant approach to construct such a constraint is by combining Maxwell's equations at both the fundamental and the second harmonic frequencies and deriving a corresponding power-conservation constraint (see section 1 in Supplement 1 for details):
\begin{equation}\label{eq:totalp}
\begin{aligned}
    &\im{\int_\Omega \vb{E}_1^{{inc}*} \cdot \vb{P}_1 \ d\vb{r}} \\
    &= \int_\Omega \vb{P}_1^* \cdot \left( \mathbb{G}_1 + \chi_1^{-\dagger}\Id \right)^a \vb{P}_1 \ d\vb{r}
    + \int_\Omega \vb{P}_2^* \cdot \mathbb{G}_2^a \vb{P}_2 \ d\vb{r} \\
    &+ \int_\Omega \left(\vb{P}_2 - \vb{P}^{(2)}\right)^* \cdot (\chi_2^{-\dagger})^a \left( \vb{P}_2 - \vb{P}^{(2)} \right) \ d\vb{r}.
\end{aligned}
\end{equation}
Here, the superscript $( ^\dagger)$ stands for conjugated transpose of an operator, $(^a)$ denotes the anti-symmetric part of an operator, $\mathbb{O}^a = (\mathbb{O}-\mathbb{O}^\dagger)/2i$, and $\mathbb{O}^{-\dagger} = \left(\mathbb{O}^{-1}\right)^\dagger$. The semi-definiteness of the anti-symmetric part of Green's function combined with the definiteness of the imaginary part of the linear responses of the material makes (\ref{eq:totalp}) a convex constraint. Physically, (\ref{eq:totalp}) is precisely the statement that power drawn from the incident field is equal to the sum of radiative and absorptive loss mechanisms at both frequencies. We note that near the dual optimum, this convex constraint, and thus, down conversion can actually be dropped without leaving the feasible domain or affecting the optimum value.
\begin{figure*}
  \includegraphics[width=\textwidth]{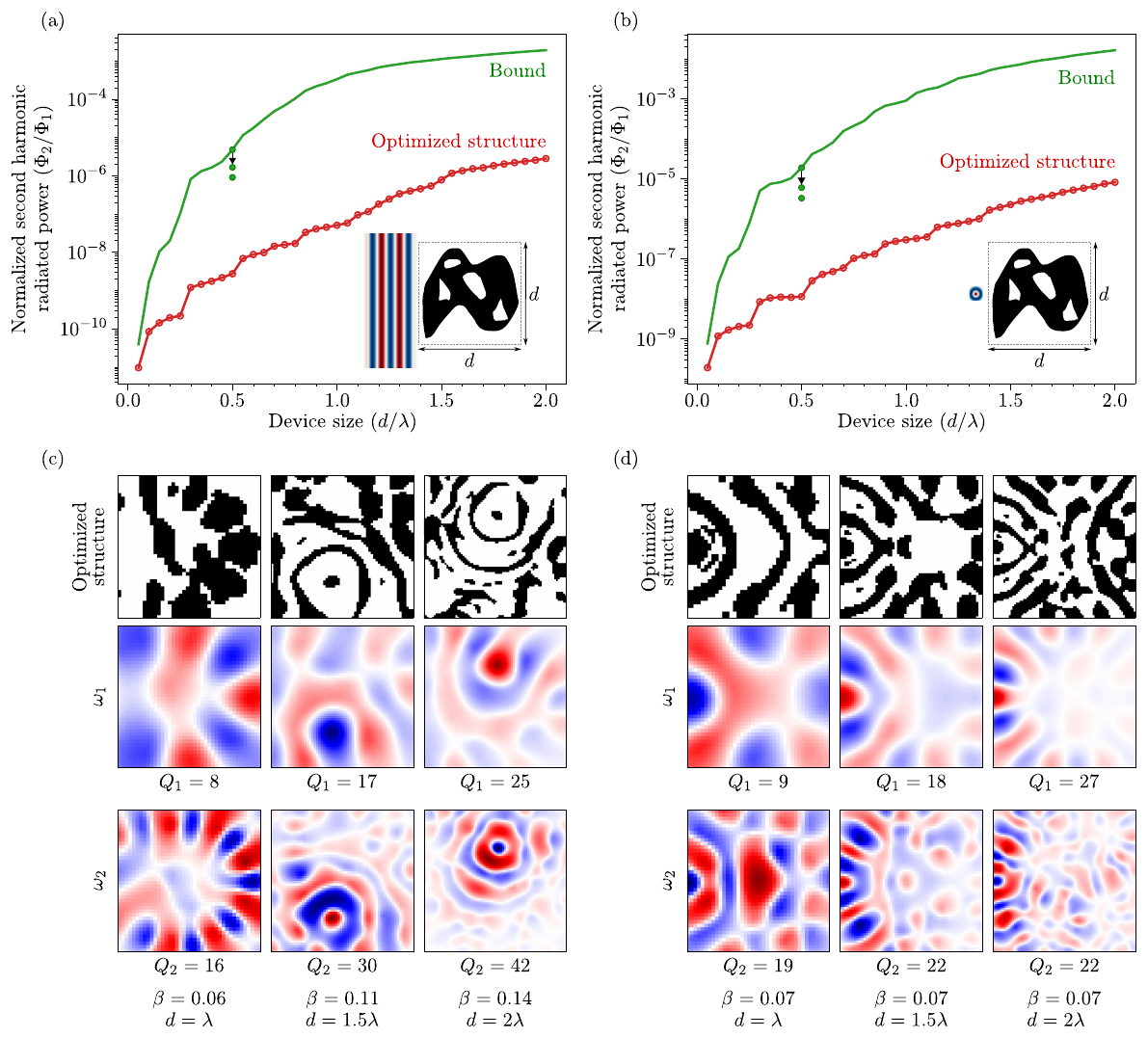}
  \caption{Upper bounds on the maximum radiative second harmonic power possible from either an (a) incident planewave or (b) dipolar current source $0.1\lambda$ away from the center of the edge of a square design region of varying size $d \times d$. The non-dispersive medium under consideration has linear susceptibility $\chi_1=\chi_2 = 3+0.1i$ and Pockels susceptibility $\chi^{(2)} = 398$ pm/V; the vacuum wavelength at $\omega_1$ is set at $\lambda = 2\pi c/\omega_1 = 1.5$ $\mu$m. The radiated power at the second harmonic frequency $\omega_2=2\omega_1$, denoted by $\Phi_2 = \frac{1}{2} \omega_2 \varepsilon_0 \im{\int_\Omega \vb{P}_2^* \cdot \mathbb{G}_2 \vb{P}_2 \ d\vb{r}}$, is normalized by either the (a) power from flux $\Phi_{1} =  \frac{1}{2}\re{\int d\vb{S}\cdot \left( \vb{E}_1^{inc} \times \vb{H}_1^{inc*} \right)} = 0.064$ mW/$\mu$m carried by the planewave incident on the largest design size considered ($d=2\lambda$), or (b) the net power emitted from the dipolar source in vacuum, $\Phi_1 = -\frac{1}{2} \re{\int \vb{J}^{inc*} \cdot \vb{E}_1^{inc} \ d\vb{r}} = 0.197$ mW/$\mu$m. All calculations were restricted to 2D fields resulting from TM-polarized (out of the plane) sources. The solid lines only incorporate global constraints across the design region $\Omega$, consistent with the choice $\{\Omega_k\}=\{\Omega_j\}=\{\Omega\}$ in \eqref{eq:opt2}; filled circles demonstrate how the imposition of additional norm constraints (\ref{eq:nlnorm}) (larger sets of $\{\Omega_j\}$) can lead to tighter bounds, here only shown for a single representative $d=0.5\lambda$. Also shown are values of $\Phi_2/\Phi_1$ achieved by optimized structures (open circles). Representative dielectric and modal electric-field profiles at both wavelengths are depicted in (c) and (d), with black/white denoting dielectric/vacuum regions and red/blue/white denoting positive/negative/zero field magnitudes. The cavity mode-overlap factor $\beta$ is given in units of $\frac{\chi^{(2)}}{\sqrt{\varepsilon_0\lambda^2}}$.} 
  \label{fig:2Dexample}
\end{figure*}

While the central derivations leading to \eqref{eq:opt2} pertain to second harmonic generation, similar procedures can be applied to formulate bounds on higher-order nonlinear processes. The reduction of higher than quadratic field dependencies to quadratic forms via the introduction of auxiliary degrees of freedom and relaxed nonlinear constraints should, in principle, be straightforward to carry out in other nonlinear settings, including Raman scattering~\cite{michon_limits_2019,langer_present_2020} and Kerr media~\cite{kippenberg_kerr-nonlinearity_2004,morin_self-kerr_2022}.
\begin{figure*}[t!]
\centering
\includegraphics[width=\linewidth]{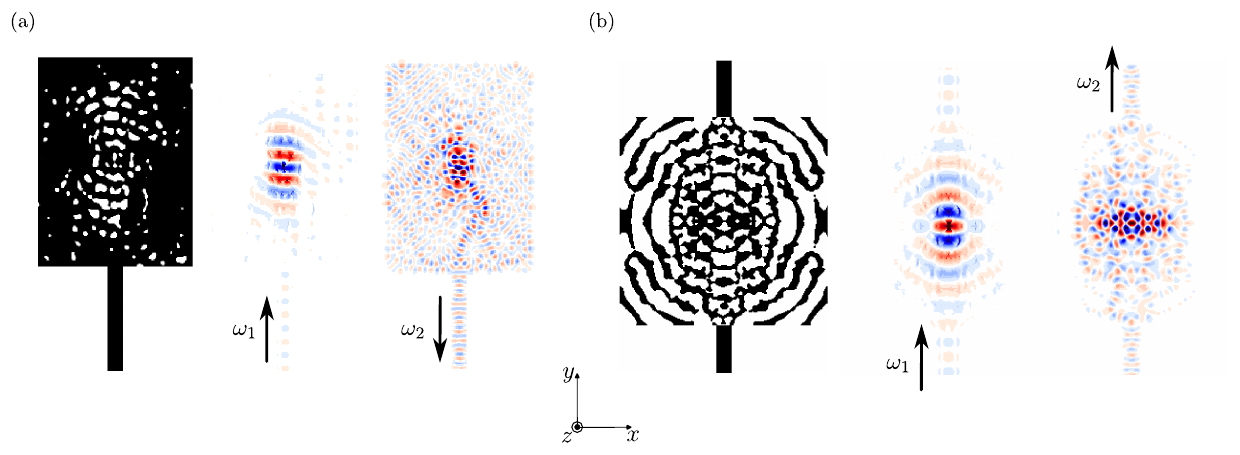}
\caption{Lateral cross-sections of 3D optimized doubly resonant cavities---gallium phosphate (GaP) structures sitting on top of silica substrates---designed for efficient second harmonic generation under critical coupling with either 
(a) a single or (b) separate input/output waveguide(s). Black/white represent GaP/vacuum regions. Device (a) has lateral size $4.65\times6.2$~$\mu$m$^2$ with a vertical thickness of 250~nm and an input waveguide of the same vertical thickness and width 457~nm; device (b) has lateral size $6.229\times6.229$~$\mu$m$^2$ with a vertical thickness of 200~nm and input/output waveguides of the same vertical thickness and width 465~nm. Both devices support resonances at the fundamental $\lambda_1=1550$~nm and second-harmonic $\lambda_2=775$~nm wavelengths with radiative quality factors (a) $Q_1 \approx 3000$ and $Q_2 \approx 1000$, and (b) $Q_1 \approx 1100$ and $Q_2 \approx 760$, respectively.  Also shown are $E_x$ and $E_z$ field profiles of the fundamental ($\omega_1$) and second harmonic ($\omega_2$) modes, respectively, with red/blue/white denoting positive/negative/zero field magnitudes.}
\label{fig:3dcoupler}
\end{figure*}

\textbf{Results---}As proof of concept, we compute bounds for two representative problems, with results shown in Fig.~\ref{fig:2Dexample}: namely, maximizing the radiative second-harmonic power produced by either (a) a normally incident planewave or (b) a dipolar source in the vicinity of a structure contained within a square design region of size $d \times d$. For computational expedience, the calculations are carried out in 2D with sources and fields polarized in the out-of-plane direction (TM polarization). We assume an isotropic, lossy medium with a weak nonlinearity and normalize resulting power quantities by either (a) the net radiation of the dipole source in vacuum or (b) the power carried by the planewave incident on the largest of the design regions. Bound calculations were performed by enforcing only the loosest of constraints, consistent with a choice of global domains $\{\Omega_k\}=\{\Omega_j\}=\{\Omega\}$ (solid line), with the addition of more localized constraints (filled circles) enforced by introducing smaller sub-domains leading to further tightening of the bounds (illustrated only for a single representative design region). See also Fig. S1 in Supplement 1 for a comparison with bounds obtained by imposing only the norm constraint in (\ref{eq:nl2}) for a global domain, $\{\Omega_j\}=\{\Omega\}$, and passivity as in~\cite{michon_limits_2019}; these are found to be many orders of magnitude looser than the bounds in Fig.~\ref{fig:2Dexample} even for materials with considerable loss. The bounds are compared to the performances of inverse designs (open circles) produced by solving the structural optimization problem of enhancing radiative SHG via gradient-based algorithms, including the method of moving asymptotes~\cite{svanberg_method_1987}.

As shown in Fig.~\ref{fig:2Dexample}, bound calculations not only place relatively tight constraints on additional improvements that may be gained from further fine-tuning of structural optimizations (coming within one or at most two orders of magnitude of inverse designs) but also anticipate several noteworthy trends seen in optimized designs with increasing system size. Both bounds and the inverse designs seem to grow polynomially for subwavelength devices with resonances appearing in optimized structures around $d=\lambda$ (vacuum wavelength), with larger devices leading to higher quality factors $Q_1$, $Q_2$. The quality factors remain relatively small due to the high material loss, which explains the performance saturation with increasing system size (in more realistic settings, such as in 3D structured slabs, radiative losses are expected to play a similar role as material loss in this 2D example, leading to decreased performance or potentially saturation, a subject for future investigations). The mode profiles of optimized structures exhibit significant nonlinear overlaps in either scenario: in the particular case of a design region of size $2\lambda$ x $2\lambda$, $\beta\approx 0.14 \frac{\chi^{(2)}}{\sqrt{\varepsilon_0 \lambda^2}}$ for the planewave source and $\beta\approx 0.07 \frac{\chi^{(2)}}{\sqrt{\varepsilon_0 \lambda^2}}$ for the corresponding dipole source. 
Inspection of inverse structures also reveals a lack of consistent structural features in designs optimized for planewave compared to dipole sources: essentially, while there are many ways to ``interfere" with a propagating wave comprising a narrow range of spatial wavelengths, dipolar fields include fast-decaying evanescent components (near fields) that require more sharply varying polarization profiles, restricting possible structures. 

While the bound calculations above focused on illustrative 2D examples, similar investigations may be carried out in more realistic settings. For example, Fig.~\ref{fig:3dcoupler} depicts optimized three-dimensional structures obtained by addressing the more challenging problem of maximizing second-harmonic power for on-chip light routed by a waveguide into a design domain. Taking advantage of the same simplifying assumptions of weak nonlinearity, undepleted pump, and bimodal approximations in (\ref{eq:opt1}), we consider maximization of the second harmonic Poynting flux, $\frac{1}{2}\re{\int_{\mathrm{wvg}} \mathrm{d}\mathbf{S}\cdot \left[ \mathbf{E}_2 \times \mathbf{H}_2^{*} \right]}$, transmitted into a prescribed output waveguide. Schematics showing lateral cross sections of the geometries are depicted in Fig.~\ref{fig:3dcoupler}, which consist of GaP on a thermal SiO$_2$ substrate as an integrated photonic platform to realize high conversion efficiency in robust, compact, and wide-bandwidth cavities~\cite{logan_400w_2018}. A single port system is considered, with either \ref{fig:3dcoupler}(a) a single or \ref{fig:3dcoupler}(b) separate waveguides guiding light at the input and output wavelengths. The $\chi^{(2)}$ tensor of GaP in the zincblende crystal phase is off-diagonal, allowing nonlinear interactions only between mutually perpendicular field components. Fig.~\ref{fig:3dcoupler} also depicts lateral cross-sections of the TE-like mode profiles (electric field mostly in-plane) at the fundamental wavelength $\lambda_1=1550$~nm and TM-like mode profiles (electric field mostly out-of-plane) at the second-harmonic wavelength $\lambda_2=775$~nm for both the structures. For ease of fabrication, the structures are designed to be invariant along the thickness dimension and to have minimal in-plane feature sizes $\gtrsim 60$~nm, leading to a vertical-to-lateral etch aspect ratio well within current experimental capabilities. Both cavities support resonances with moderate radiative quality factors $\sim 10^3$ and tightly confined modes, leading to large spatial overlaps, $\beta \approx 0.01 \frac{\chi^{(2)}}{\sqrt{\varepsilon_0 \lambda^3}}$, while ensuring near-critical coupling to nearby input/output waveguides.

\textbf{Conclusion---}We presented an approach to establish fundamental limits on nonlinear photonics objectives and applied it to investigate second harmonic generation in wavelength-scale structures. The limits not only quantify potential room for improvements but may be used to study scaling characteristics of optimal devices with respect to elemental design criteria, such as material choice and device size, without reference to particular geometric or physical enhancement mechanisms. A thorough and more focused exploration of limits on SHG in more realistic settings, such as the slab geometry of Fig.~\ref{fig:3dcoupler}, remains a challenging but important problem for future work. In addition to overcoming computational challenges needed to incorporate greater numbers of more finely resolved spatial constraints to obtain tighter predictions, there are several interesting directions worth pursuing. First, it should be possible to exploit symmetries to enforce that bounds respect fabrication constraints, including minimum-feature sizes and etchable geometries. Second, extension of the proposed framework to incorporate finite-bandwidth objectives should enable analysis of nonlinear space-bandwidth limitations, including trade-offs in achieving maximum spatial confinement and operating speeds at both wavelengths. Finally, detailed comparisons with established geometries like ring resonators~\cite{zhang_monolithic_2017,lu_periodically_2019,lu_toward_2020} and inverse designs, including performance comparisons of dielectric, metallic, or even heterogeneous~\cite{amaolo_performance_2023} media, should provide a more comprehensive view of the optimal design and performance landscape.

We acknowledge the support by the National Science Foundation under the Emerging Frontiers in Research and Innovation (EFRI) program, Award No. EFMA164098, the Defense Advanced Research Projects Agency (DARPA) under Agreements No. HR00111820046, No. HR00112090011, and No. HR0011047197, and by a Princeton SEAS Innovation Grant. SM acknowledges financial support from IVADO (Institut de valorisation des données, Québec). The simulations presented in this article were performed on computational resources managed and supported by Princeton Research Computing, a consortium of groups including the Princeton Institute for Computational Science and Engineering (PICSciE) and the Office of Information Technology’s High Performance Computing Center and Visualization Laboratory at Princeton University. The views, opinions, and findings expressed herein are those of the authors and should not be interpreted as representing the official views or policies of any institution.

\bibliography{shg_refs}

\begin{thebibliography}{36}
\expandafter\ifx\csname natexlab\endcsname\relax\def\natexlab#1{#1}\fi
\expandafter\ifx\csname bibnamefont\endcsname\relax
  \def\bibnamefont#1{#1}\fi
\expandafter\ifx\csname bibfnamefont\endcsname\relax
  \def\bibfnamefont#1{#1}\fi
\expandafter\ifx\csname citenamefont\endcsname\relax
  \def\citenamefont#1{#1}\fi
\expandafter\ifx\csname url\endcsname\relax
  \def\url#1{\texttt{#1}}\fi
\expandafter\ifx\csname urlprefix\endcsname\relax\def\urlprefix{URL }\fi
\providecommand{\bibinfo}[2]{#2}
\providecommand{\eprint}[2][]{\url{#2}}

\bibitem[{\citenamefont{Chao et~al.}(2022{\natexlab{a}})\citenamefont{Chao,
  Strekha, Kuate~Defo, Molesky, and Rodriguez}}]{chao_physical_2022}
\bibinfo{author}{\bibfnamefont{P.}~\bibnamefont{Chao}},
  \bibinfo{author}{\bibfnamefont{B.}~\bibnamefont{Strekha}},
  \bibinfo{author}{\bibfnamefont{R.}~\bibnamefont{Kuate~Defo}},
  \bibinfo{author}{\bibfnamefont{S.}~\bibnamefont{Molesky}}, \bibnamefont{and}
  \bibinfo{author}{\bibfnamefont{A.~W.} \bibnamefont{Rodriguez}},
  \bibinfo{journal}{Nature Reviews Physics} \textbf{\bibinfo{volume}{4}},
  \bibinfo{pages}{543} (\bibinfo{year}{2022}{\natexlab{a}}), ISSN
  \bibinfo{issn}{2522-5820}.

\bibitem[{\citenamefont{Angeris et~al.}(2021)\citenamefont{Angeris, Vu{\v
  c}kovi{\'c}, and Boyd}}]{angeris_heuristic_2021}
\bibinfo{author}{\bibfnamefont{G.}~\bibnamefont{Angeris}},
  \bibinfo{author}{\bibfnamefont{J.}~\bibnamefont{Vu{\v c}kovi{\'c}}},
  \bibnamefont{and} \bibinfo{author}{\bibfnamefont{S.}~\bibnamefont{Boyd}},
  \bibinfo{journal}{Optics Express} \textbf{\bibinfo{volume}{29}},
  \bibinfo{pages}{2827} (\bibinfo{year}{2021}).

\bibitem[{\citenamefont{Gertler et~al.}(2023)\citenamefont{Gertler, Kuang,
  Christie, and Miller}}]{gertler_many_2023}
\bibinfo{author}{\bibfnamefont{S.}~\bibnamefont{Gertler}},
  \bibinfo{author}{\bibfnamefont{Z.}~\bibnamefont{Kuang}},
  \bibinfo{author}{\bibfnamefont{C.}~\bibnamefont{Christie}}, \bibnamefont{and}
  \bibinfo{author}{\bibfnamefont{O.~D.} \bibnamefont{Miller}},
  \emph{\bibinfo{title}{Many {{Physical Design Problems}} are {{Sparse
  QCQPs}}}} (\bibinfo{year}{2023}), \eprint{2303.17691}.

\bibitem[{\citenamefont{Gustafsson et~al.}(2020)\citenamefont{Gustafsson,
  Schab, Jelinek, and Capek}}]{gustafsson_upper_2020}
\bibinfo{author}{\bibfnamefont{M.}~\bibnamefont{Gustafsson}},
  \bibinfo{author}{\bibfnamefont{K.}~\bibnamefont{Schab}},
  \bibinfo{author}{\bibfnamefont{L.}~\bibnamefont{Jelinek}}, \bibnamefont{and}
  \bibinfo{author}{\bibfnamefont{M.}~\bibnamefont{Capek}},
  \bibinfo{journal}{New Journal of Physics} \textbf{\bibinfo{volume}{22}},
  \bibinfo{pages}{073013} (\bibinfo{year}{2020}), ISSN
  \bibinfo{issn}{1367-2630}.

\bibitem[{\citenamefont{Chen et~al.}(1995)\citenamefont{Chen, Wang, Wu, Wu,
  Zeng, and Yu}}]{chen_design_1995}
\bibinfo{author}{\bibfnamefont{C.}~\bibnamefont{Chen}},
  \bibinfo{author}{\bibfnamefont{Y.}~\bibnamefont{Wang}},
  \bibinfo{author}{\bibfnamefont{B.}~\bibnamefont{Wu}},
  \bibinfo{author}{\bibfnamefont{K.}~\bibnamefont{Wu}},
  \bibinfo{author}{\bibfnamefont{W.}~\bibnamefont{Zeng}}, \bibnamefont{and}
  \bibinfo{author}{\bibfnamefont{L.}~\bibnamefont{Yu}},
  \bibinfo{journal}{Nature} \textbf{\bibinfo{volume}{373}},
  \bibinfo{pages}{322} (\bibinfo{year}{1995}), ISSN \bibinfo{issn}{1476-4687}.

\bibitem[{\citenamefont{Chen et~al.}(2021)\citenamefont{Chen, Hu, Kong, and
  Mao}}]{chen_high-performance_2021}
\bibinfo{author}{\bibfnamefont{J.}~\bibnamefont{Chen}},
  \bibinfo{author}{\bibfnamefont{C.-L.} \bibnamefont{Hu}},
  \bibinfo{author}{\bibfnamefont{F.}~\bibnamefont{Kong}}, \bibnamefont{and}
  \bibinfo{author}{\bibfnamefont{J.-G.} \bibnamefont{Mao}},
  \bibinfo{journal}{Accounts of Chemical Research}
  \textbf{\bibinfo{volume}{54}}, \bibinfo{pages}{2775} (\bibinfo{year}{2021}),
  ISSN \bibinfo{issn}{0001-4842}.

\bibitem[{\citenamefont{Liu et~al.}(2015)\citenamefont{Liu, Ye, Khan, and
  Sorger}}]{liu_review_2015}
\bibinfo{author}{\bibfnamefont{K.}~\bibnamefont{Liu}},
  \bibinfo{author}{\bibfnamefont{C.~R.} \bibnamefont{Ye}},
  \bibinfo{author}{\bibfnamefont{S.}~\bibnamefont{Khan}}, \bibnamefont{and}
  \bibinfo{author}{\bibfnamefont{V.~J.} \bibnamefont{Sorger}},
  \bibinfo{journal}{Laser \& Photonics Reviews} \textbf{\bibinfo{volume}{9}},
  \bibinfo{pages}{172} (\bibinfo{year}{2015}), ISSN \bibinfo{issn}{1863-8899}.

\bibitem[{\citenamefont{Pavone and Campagnola}(2014)}]{pavone_second_2014}
\bibinfo{editor}{\bibfnamefont{F.~S.} \bibnamefont{Pavone}} \bibnamefont{and}
  \bibinfo{editor}{\bibfnamefont{P.~J.} \bibnamefont{Campagnola}}, eds.,
  \emph{\bibinfo{title}{Second Harmonic Generation Imaging}},
  no.~\bibinfo{number}{3} in \bibinfo{series}{Series in Cellular and Clinical
  Imaging} (\bibinfo{publisher}{{CRC Press Taylor \& Francis}},
  \bibinfo{address}{{Boca Raton}}, \bibinfo{year}{2014}), ISBN
  \bibinfo{isbn}{978-1-4398-4914-9}.

\bibitem[{\citenamefont{Heinz et~al.}(1982)\citenamefont{Heinz, Chen, Ricard,
  and Shen}}]{heinz_spectroscopy_1982}
\bibinfo{author}{\bibfnamefont{T.~F.} \bibnamefont{Heinz}},
  \bibinfo{author}{\bibfnamefont{C.~K.} \bibnamefont{Chen}},
  \bibinfo{author}{\bibfnamefont{D.}~\bibnamefont{Ricard}}, \bibnamefont{and}
  \bibinfo{author}{\bibfnamefont{Y.~R.} \bibnamefont{Shen}},
  \bibinfo{journal}{Physical Review Letters} \textbf{\bibinfo{volume}{48}},
  \bibinfo{pages}{478} (\bibinfo{year}{1982}).

\bibitem[{\citenamefont{Wang et~al.}(2019)\citenamefont{Wang, Xiao, Yang, Wang,
  and Zhang}}]{wang_second_2019}
\bibinfo{author}{\bibfnamefont{Y.}~\bibnamefont{Wang}},
  \bibinfo{author}{\bibfnamefont{J.}~\bibnamefont{Xiao}},
  \bibinfo{author}{\bibfnamefont{S.}~\bibnamefont{Yang}},
  \bibinfo{author}{\bibfnamefont{Y.}~\bibnamefont{Wang}}, \bibnamefont{and}
  \bibinfo{author}{\bibfnamefont{X.}~\bibnamefont{Zhang}},
  \bibinfo{journal}{Optical Materials Express} \textbf{\bibinfo{volume}{9}},
  \bibinfo{pages}{1136} (\bibinfo{year}{2019}), ISSN \bibinfo{issn}{2159-3930}.

\bibitem[{\citenamefont{Lin et~al.}(2016)\citenamefont{Lin, Liang, Lon{\v c}ar,
  Johnson, and Rodriguez}}]{lin_cavity-enhanced_2016}
\bibinfo{author}{\bibfnamefont{Z.}~\bibnamefont{Lin}},
  \bibinfo{author}{\bibfnamefont{X.}~\bibnamefont{Liang}},
  \bibinfo{author}{\bibfnamefont{M.}~\bibnamefont{Lon{\v c}ar}},
  \bibinfo{author}{\bibfnamefont{S.~G.} \bibnamefont{Johnson}},
  \bibnamefont{and} \bibinfo{author}{\bibfnamefont{A.~W.}
  \bibnamefont{Rodriguez}}, \bibinfo{journal}{Optica}
  \textbf{\bibinfo{volume}{3}}, \bibinfo{pages}{233} (\bibinfo{year}{2016}).

\bibitem[{\citenamefont{Rodriguez et~al.}(2007)\citenamefont{Rodriguez,
  Solja{\v c}i{\'c}, Joannopoulos, and Johnson}}]{rodriguez_2_2007}
\bibinfo{author}{\bibfnamefont{A.}~\bibnamefont{Rodriguez}},
  \bibinfo{author}{\bibfnamefont{M.}~\bibnamefont{Solja{\v c}i{\'c}}},
  \bibinfo{author}{\bibfnamefont{J.~D.} \bibnamefont{Joannopoulos}},
  \bibnamefont{and} \bibinfo{author}{\bibfnamefont{S.~G.}
  \bibnamefont{Johnson}}, \bibinfo{journal}{Optics Express}
  \textbf{\bibinfo{volume}{15}}, \bibinfo{pages}{7303} (\bibinfo{year}{2007}),
  ISSN \bibinfo{issn}{1094-4087}.

\bibitem[{\citenamefont{Chao et~al.}(2022{\natexlab{b}})\citenamefont{Chao,
  Defo, Molesky, and Rodriguez}}]{chao_maximum_2022}
\bibinfo{author}{\bibfnamefont{P.}~\bibnamefont{Chao}},
  \bibinfo{author}{\bibfnamefont{R.~K.} \bibnamefont{Defo}},
  \bibinfo{author}{\bibfnamefont{S.}~\bibnamefont{Molesky}}, \bibnamefont{and}
  \bibinfo{author}{\bibfnamefont{A.}~\bibnamefont{Rodriguez}},
  \bibinfo{journal}{Nanophotonics}  (\bibinfo{year}{2022}{\natexlab{b}}), ISSN
  \bibinfo{issn}{2192-8614}.

\bibitem[{\citenamefont{Joannopoulos et~al.}(2008)\citenamefont{Joannopoulos,
  Steven, Winn, and Meade}}]{joannopoulos_photonic_2008}
\bibinfo{author}{\bibfnamefont{J.~D.} \bibnamefont{Joannopoulos}},
  \bibinfo{author}{\bibfnamefont{J.~G.} \bibnamefont{Steven}},
  \bibinfo{author}{\bibfnamefont{J.~N.} \bibnamefont{Winn}}, \bibnamefont{and}
  \bibinfo{author}{\bibfnamefont{R.~D.} \bibnamefont{Meade}},
  \emph{\bibinfo{title}{Photonic Crystals: Molding the Flow of Light}}
  (\bibinfo{publisher}{{Princeton University Press}},
  \bibinfo{address}{{Princeton}}, \bibinfo{year}{2008}), \bibinfo{edition}{2nd}
  ed., ISBN \bibinfo{isbn}{978-0-691-12456-8}.

\bibitem[{\citenamefont{Fejer}(1994)}]{fejer_nonlinear_1994}
\bibinfo{author}{\bibfnamefont{M.~M.} \bibnamefont{Fejer}},
  \bibinfo{journal}{Physics Today} \textbf{\bibinfo{volume}{47}},
  \bibinfo{pages}{25} (\bibinfo{year}{1994}), ISSN \bibinfo{issn}{0031-9228}.

\bibitem[{\citenamefont{F{\"u}rst et~al.}(2010)\citenamefont{F{\"u}rst,
  Strekalov, Elser, Lassen, Andersen, Marquardt, and
  Leuchs}}]{furst_naturally_2010}
\bibinfo{author}{\bibfnamefont{J.~U.} \bibnamefont{F{\"u}rst}},
  \bibinfo{author}{\bibfnamefont{D.~V.} \bibnamefont{Strekalov}},
  \bibinfo{author}{\bibfnamefont{D.}~\bibnamefont{Elser}},
  \bibinfo{author}{\bibfnamefont{M.}~\bibnamefont{Lassen}},
  \bibinfo{author}{\bibfnamefont{U.~L.} \bibnamefont{Andersen}},
  \bibinfo{author}{\bibfnamefont{C.}~\bibnamefont{Marquardt}},
  \bibnamefont{and} \bibinfo{author}{\bibfnamefont{G.}~\bibnamefont{Leuchs}},
  \bibinfo{journal}{Physical Review Letters} \textbf{\bibinfo{volume}{104}},
  \bibinfo{pages}{153901} (\bibinfo{year}{2010}).

\bibitem[{\citenamefont{Bi et~al.}(2012)\citenamefont{Bi, Rodriguez, Hashemi,
  Duchesne, Loncar, Wang, and Johnson}}]{bi_high-efficiency_2012}
\bibinfo{author}{\bibfnamefont{Z.-F.} \bibnamefont{Bi}},
  \bibinfo{author}{\bibfnamefont{A.~W.} \bibnamefont{Rodriguez}},
  \bibinfo{author}{\bibfnamefont{H.}~\bibnamefont{Hashemi}},
  \bibinfo{author}{\bibfnamefont{D.}~\bibnamefont{Duchesne}},
  \bibinfo{author}{\bibfnamefont{M.}~\bibnamefont{Loncar}},
  \bibinfo{author}{\bibfnamefont{K.-M.} \bibnamefont{Wang}}, \bibnamefont{and}
  \bibinfo{author}{\bibfnamefont{S.~G.} \bibnamefont{Johnson}},
  \bibinfo{journal}{Optics Express} \textbf{\bibinfo{volume}{20}},
  \bibinfo{pages}{7526} (\bibinfo{year}{2012}), ISSN \bibinfo{issn}{1094-4087}.

\bibitem[{\citenamefont{Logan et~al.}(2018)\citenamefont{Logan, Gould,
  Schmidgall, Hestroffer, Lin, Jin, Majumdar, Hatami, Rodriguez, and
  Fu}}]{logan_400w_2018}
\bibinfo{author}{\bibfnamefont{A.~D.} \bibnamefont{Logan}},
  \bibinfo{author}{\bibfnamefont{M.}~\bibnamefont{Gould}},
  \bibinfo{author}{\bibfnamefont{E.~R.} \bibnamefont{Schmidgall}},
  \bibinfo{author}{\bibfnamefont{K.}~\bibnamefont{Hestroffer}},
  \bibinfo{author}{\bibfnamefont{Z.}~\bibnamefont{Lin}},
  \bibinfo{author}{\bibfnamefont{W.}~\bibnamefont{Jin}},
  \bibinfo{author}{\bibfnamefont{A.}~\bibnamefont{Majumdar}},
  \bibinfo{author}{\bibfnamefont{F.}~\bibnamefont{Hatami}},
  \bibinfo{author}{\bibfnamefont{A.~W.} \bibnamefont{Rodriguez}},
  \bibnamefont{and} \bibinfo{author}{\bibfnamefont{K.-M.~C.} \bibnamefont{Fu}},
  \bibinfo{journal}{Optics Express} \textbf{\bibinfo{volume}{26}},
  \bibinfo{pages}{33687} (\bibinfo{year}{2018}), ISSN
  \bibinfo{issn}{1094-4087}.

\bibitem[{\citenamefont{Pernice et~al.}(2012)\citenamefont{Pernice, Xiong,
  Schuck, and Tang}}]{pernice_second_2012}
\bibinfo{author}{\bibfnamefont{W.~H.~P.} \bibnamefont{Pernice}},
  \bibinfo{author}{\bibfnamefont{C.}~\bibnamefont{Xiong}},
  \bibinfo{author}{\bibfnamefont{C.}~\bibnamefont{Schuck}}, \bibnamefont{and}
  \bibinfo{author}{\bibfnamefont{H.~X.} \bibnamefont{Tang}},
  \bibinfo{journal}{Applied Physics Letters} \textbf{\bibinfo{volume}{100}},
  \bibinfo{pages}{223501} (\bibinfo{year}{2012}), ISSN
  \bibinfo{issn}{0003-6951}.

\bibitem[{\citenamefont{{Bravo-Abad} et~al.}(2010)\citenamefont{{Bravo-Abad},
  Rodriguez, Joannopoulos, Rakich, Johnson, and Solja{\v
  c}i{\'c}}}]{bravo-abad_efficient_2010}
\bibinfo{author}{\bibfnamefont{J.}~\bibnamefont{{Bravo-Abad}}},
  \bibinfo{author}{\bibfnamefont{A.~W.} \bibnamefont{Rodriguez}},
  \bibinfo{author}{\bibfnamefont{J.~D.} \bibnamefont{Joannopoulos}},
  \bibinfo{author}{\bibfnamefont{P.~T.} \bibnamefont{Rakich}},
  \bibinfo{author}{\bibfnamefont{S.~G.} \bibnamefont{Johnson}},
  \bibnamefont{and} \bibinfo{author}{\bibfnamefont{M.}~\bibnamefont{Solja{\v
  c}i{\'c}}}, \bibinfo{journal}{Applied Physics Letters}
  \textbf{\bibinfo{volume}{96}}, \bibinfo{pages}{101110}
  (\bibinfo{year}{2010}), ISSN \bibinfo{issn}{0003-6951}.

\bibitem[{\citenamefont{Michon et~al.}(2019)\citenamefont{Michon, Benzaouia,
  Yao, Miller, and Johnson}}]{michon_limits_2019}
\bibinfo{author}{\bibfnamefont{J.}~\bibnamefont{Michon}},
  \bibinfo{author}{\bibfnamefont{M.}~\bibnamefont{Benzaouia}},
  \bibinfo{author}{\bibfnamefont{W.}~\bibnamefont{Yao}},
  \bibinfo{author}{\bibfnamefont{O.~D.} \bibnamefont{Miller}},
  \bibnamefont{and} \bibinfo{author}{\bibfnamefont{S.~G.}
  \bibnamefont{Johnson}}, \bibinfo{journal}{Optics Express}
  \textbf{\bibinfo{volume}{27}}, \bibinfo{pages}{35189} (\bibinfo{year}{2019}).

\bibitem[{\citenamefont{Miller et~al.}(2016)\citenamefont{Miller, Polimeridis,
  Reid, Hsu, DeLacy, Joannopoulos, Solja{\v c}i{\'c}, and
  Johnson}}]{miller_fundamental_2016}
\bibinfo{author}{\bibfnamefont{O.~D.} \bibnamefont{Miller}},
  \bibinfo{author}{\bibfnamefont{A.~G.} \bibnamefont{Polimeridis}},
  \bibinfo{author}{\bibfnamefont{M.~T.~H.} \bibnamefont{Reid}},
  \bibinfo{author}{\bibfnamefont{C.~W.} \bibnamefont{Hsu}},
  \bibinfo{author}{\bibfnamefont{B.~G.} \bibnamefont{DeLacy}},
  \bibinfo{author}{\bibfnamefont{J.~D.} \bibnamefont{Joannopoulos}},
  \bibinfo{author}{\bibfnamefont{M.}~\bibnamefont{Solja{\v c}i{\'c}}},
  \bibnamefont{and} \bibinfo{author}{\bibfnamefont{S.~G.}
  \bibnamefont{Johnson}}, \bibinfo{journal}{Optics Express}
  \textbf{\bibinfo{volume}{24}}, \bibinfo{pages}{3329} (\bibinfo{year}{2016}),
  ISSN \bibinfo{issn}{1094-4087}.

\bibitem[{\citenamefont{Molesky et~al.}(2022)\citenamefont{Molesky, Chao,
  Mohajan, Reinhart, Chi, and Rodriguez}}]{molesky_t_2022}
\bibinfo{author}{\bibfnamefont{S.}~\bibnamefont{Molesky}},
  \bibinfo{author}{\bibfnamefont{P.}~\bibnamefont{Chao}},
  \bibinfo{author}{\bibfnamefont{J.}~\bibnamefont{Mohajan}},
  \bibinfo{author}{\bibfnamefont{W.}~\bibnamefont{Reinhart}},
  \bibinfo{author}{\bibfnamefont{H.}~\bibnamefont{Chi}}, \bibnamefont{and}
  \bibinfo{author}{\bibfnamefont{A.~W.} \bibnamefont{Rodriguez}},
  \bibinfo{journal}{Physical Review Research} \textbf{\bibinfo{volume}{4}},
  \bibinfo{pages}{013020} (\bibinfo{year}{2022}), ISSN
  \bibinfo{issn}{2643-1564}.

\bibitem[{\citenamefont{Molesky
  et~al.}(2020{\natexlab{a}})\citenamefont{Molesky, Chao, Jin, and
  Rodriguez}}]{molesky_global_2020}
\bibinfo{author}{\bibfnamefont{S.}~\bibnamefont{Molesky}},
  \bibinfo{author}{\bibfnamefont{P.}~\bibnamefont{Chao}},
  \bibinfo{author}{\bibfnamefont{W.}~\bibnamefont{Jin}}, \bibnamefont{and}
  \bibinfo{author}{\bibfnamefont{A.~W.} \bibnamefont{Rodriguez}},
  \bibinfo{journal}{Physical Review Research} \textbf{\bibinfo{volume}{2}},
  \bibinfo{pages}{033172} (\bibinfo{year}{2020}{\natexlab{a}}).

\bibitem[{\citenamefont{Boyd}(2008)}]{boyd_nonlinear_2008}
\bibinfo{author}{\bibfnamefont{R.~W.} \bibnamefont{Boyd}},
  \emph{\bibinfo{title}{Nonlinear Optics}} (\bibinfo{publisher}{{Academic
  Press}}, \bibinfo{address}{{Amsterdam ; Boston}}, \bibinfo{year}{2008}),
  \bibinfo{edition}{3rd} ed., ISBN \bibinfo{isbn}{978-0-12-369470-6}.

\bibitem[{\citenamefont{Molesky
  et~al.}(2020{\natexlab{b}})\citenamefont{Molesky, Chao, and
  Rodriguez}}]{molesky_hierarchical_2020}
\bibinfo{author}{\bibfnamefont{S.}~\bibnamefont{Molesky}},
  \bibinfo{author}{\bibfnamefont{P.}~\bibnamefont{Chao}}, \bibnamefont{and}
  \bibinfo{author}{\bibfnamefont{A.~W.} \bibnamefont{Rodriguez}},
  \bibinfo{journal}{Physical Review Research} \textbf{\bibinfo{volume}{2}},
  \bibinfo{pages}{043398} (\bibinfo{year}{2020}{\natexlab{b}}).

\bibitem[{\citenamefont{Kuang and Miller}(2020)}]{kuang_computational_2020}
\bibinfo{author}{\bibfnamefont{Z.}~\bibnamefont{Kuang}} \bibnamefont{and}
  \bibinfo{author}{\bibfnamefont{O.~D.} \bibnamefont{Miller}},
  \bibinfo{journal}{Physical Review Letters} \textbf{\bibinfo{volume}{125}},
  \bibinfo{pages}{263607} (\bibinfo{year}{2020}).

\bibitem[{\citenamefont{Shim et~al.}(2021)\citenamefont{Shim, Kuang, Lin, and
  Miller}}]{shim_fundamental_2021}
\bibinfo{author}{\bibfnamefont{H.}~\bibnamefont{Shim}},
  \bibinfo{author}{\bibfnamefont{Z.}~\bibnamefont{Kuang}},
  \bibinfo{author}{\bibfnamefont{Z.}~\bibnamefont{Lin}}, \bibnamefont{and}
  \bibinfo{author}{\bibfnamefont{O.~D.} \bibnamefont{Miller}},
  \emph{\bibinfo{title}{Fundamental limits to multi-functional and tunable
  nanophotonic response}} (\bibinfo{year}{2021}), \eprint{2112.10816}.

\bibitem[{\citenamefont{Langer et~al.}(2020)\citenamefont{Langer, {Jimenez de
  Aberasturi}, Aizpurua, {Alvarez-Puebla}, Augui{\'e}, Baumberg, Bazan, Bell,
  Boisen, Brolo et~al.}}]{langer_present_2020}
\bibinfo{author}{\bibfnamefont{J.}~\bibnamefont{Langer}},
  \bibinfo{author}{\bibfnamefont{D.}~\bibnamefont{{Jimenez de Aberasturi}}},
  \bibinfo{author}{\bibfnamefont{J.}~\bibnamefont{Aizpurua}},
  \bibinfo{author}{\bibfnamefont{R.~A.} \bibnamefont{{Alvarez-Puebla}}},
  \bibinfo{author}{\bibfnamefont{B.}~\bibnamefont{Augui{\'e}}},
  \bibinfo{author}{\bibfnamefont{J.~J.} \bibnamefont{Baumberg}},
  \bibinfo{author}{\bibfnamefont{G.~C.} \bibnamefont{Bazan}},
  \bibinfo{author}{\bibfnamefont{S.~E.~J.} \bibnamefont{Bell}},
  \bibinfo{author}{\bibfnamefont{A.}~\bibnamefont{Boisen}},
  \bibinfo{author}{\bibfnamefont{A.~G.} \bibnamefont{Brolo}},
  \bibnamefont{et~al.}, \bibinfo{journal}{ACS Nano}
  \textbf{\bibinfo{volume}{14}}, \bibinfo{pages}{28} (\bibinfo{year}{2020}),
  ISSN \bibinfo{issn}{1936-0851}.

\bibitem[{\citenamefont{Kippenberg et~al.}(2004)\citenamefont{Kippenberg,
  Spillane, and Vahala}}]{kippenberg_kerr-nonlinearity_2004}
\bibinfo{author}{\bibfnamefont{T.~J.} \bibnamefont{Kippenberg}},
  \bibinfo{author}{\bibfnamefont{S.~M.} \bibnamefont{Spillane}},
  \bibnamefont{and} \bibinfo{author}{\bibfnamefont{K.~J.}
  \bibnamefont{Vahala}}, \bibinfo{journal}{Physical Review Letters}
  \textbf{\bibinfo{volume}{93}}, \bibinfo{pages}{083904}
  (\bibinfo{year}{2004}), ISSN \bibinfo{issn}{0031-9007, 1079-7114}.

\bibitem[{\citenamefont{Morin et~al.}(2022)\citenamefont{Morin, Tignon,
  Mangeney, Dhillon, Czajkowski, Karpi{\'n}ski, {Zieli{\'n}ska-Raczy{\'n}ska},
  Ziemkiewicz, and Boulier}}]{morin_self-kerr_2022}
\bibinfo{author}{\bibfnamefont{C.}~\bibnamefont{Morin}},
  \bibinfo{author}{\bibfnamefont{J.}~\bibnamefont{Tignon}},
  \bibinfo{author}{\bibfnamefont{J.}~\bibnamefont{Mangeney}},
  \bibinfo{author}{\bibfnamefont{S.}~\bibnamefont{Dhillon}},
  \bibinfo{author}{\bibfnamefont{G.}~\bibnamefont{Czajkowski}},
  \bibinfo{author}{\bibfnamefont{K.}~\bibnamefont{Karpi{\'n}ski}},
  \bibinfo{author}{\bibfnamefont{S.}~\bibnamefont{{Zieli{\'n}ska-Raczy{\'n}ska}}},
  \bibinfo{author}{\bibfnamefont{D.}~\bibnamefont{Ziemkiewicz}},
  \bibnamefont{and} \bibinfo{author}{\bibfnamefont{T.}~\bibnamefont{Boulier}},
  \bibinfo{journal}{Physical Review Letters} \textbf{\bibinfo{volume}{129}},
  \bibinfo{pages}{137401} (\bibinfo{year}{2022}).

\bibitem[{\citenamefont{Svanberg}(1987)}]{svanberg_method_1987}
\bibinfo{author}{\bibfnamefont{K.}~\bibnamefont{Svanberg}},
  \bibinfo{journal}{International Journal for Numerical Methods in Engineering}
  \textbf{\bibinfo{volume}{24}}, \bibinfo{pages}{359} (\bibinfo{year}{1987}),
  ISSN \bibinfo{issn}{1097-0207}.

\bibitem[{\citenamefont{Zhang et~al.}(2017)\citenamefont{Zhang, Wang, Cheng,
  {Shams-Ansari}, and Lon{\v c}ar}}]{zhang_monolithic_2017}
\bibinfo{author}{\bibfnamefont{M.}~\bibnamefont{Zhang}},
  \bibinfo{author}{\bibfnamefont{C.}~\bibnamefont{Wang}},
  \bibinfo{author}{\bibfnamefont{R.}~\bibnamefont{Cheng}},
  \bibinfo{author}{\bibfnamefont{A.}~\bibnamefont{{Shams-Ansari}}},
  \bibnamefont{and} \bibinfo{author}{\bibfnamefont{M.}~\bibnamefont{Lon{\v
  c}ar}}, \bibinfo{journal}{Optica} \textbf{\bibinfo{volume}{4}},
  \bibinfo{pages}{1536} (\bibinfo{year}{2017}), ISSN \bibinfo{issn}{2334-2536}.

\bibitem[{\citenamefont{Lu et~al.}(2019)\citenamefont{Lu, Surya, Liu, Bruch,
  Gong, Xu, and Tang}}]{lu_periodically_2019}
\bibinfo{author}{\bibfnamefont{J.}~\bibnamefont{Lu}},
  \bibinfo{author}{\bibfnamefont{J.~B.} \bibnamefont{Surya}},
  \bibinfo{author}{\bibfnamefont{X.}~\bibnamefont{Liu}},
  \bibinfo{author}{\bibfnamefont{A.~W.} \bibnamefont{Bruch}},
  \bibinfo{author}{\bibfnamefont{Z.}~\bibnamefont{Gong}},
  \bibinfo{author}{\bibfnamefont{Y.}~\bibnamefont{Xu}}, \bibnamefont{and}
  \bibinfo{author}{\bibfnamefont{H.~X.} \bibnamefont{Tang}},
  \bibinfo{journal}{Optica} \textbf{\bibinfo{volume}{6}}, \bibinfo{pages}{1455}
  (\bibinfo{year}{2019}), ISSN \bibinfo{issn}{2334-2536}.

\bibitem[{\citenamefont{Lu et~al.}(2020)\citenamefont{Lu, Li, Zou, Sayem, and
  Tang}}]{lu_toward_2020}
\bibinfo{author}{\bibfnamefont{J.}~\bibnamefont{Lu}},
  \bibinfo{author}{\bibfnamefont{M.}~\bibnamefont{Li}},
  \bibinfo{author}{\bibfnamefont{C.-L.} \bibnamefont{Zou}},
  \bibinfo{author}{\bibfnamefont{A.~A.} \bibnamefont{Sayem}}, \bibnamefont{and}
  \bibinfo{author}{\bibfnamefont{H.~X.} \bibnamefont{Tang}},
  \bibinfo{journal}{Optica} \textbf{\bibinfo{volume}{7}}, \bibinfo{pages}{1654}
  (\bibinfo{year}{2020}), ISSN \bibinfo{issn}{2334-2536}.

\bibitem[{\citenamefont{Amaolo et~al.}(2023)\citenamefont{Amaolo, Chao,
  Maldonado, Molesky, and Rodriguez}}]{amaolo_performance_2023}
\bibinfo{author}{\bibfnamefont{A.}~\bibnamefont{Amaolo}},
  \bibinfo{author}{\bibfnamefont{P.}~\bibnamefont{Chao}},
  \bibinfo{author}{\bibfnamefont{T.~J.} \bibnamefont{Maldonado}},
  \bibinfo{author}{\bibfnamefont{S.}~\bibnamefont{Molesky}}, \bibnamefont{and}
  \bibinfo{author}{\bibfnamefont{A.~W.} \bibnamefont{Rodriguez}},
  \emph{\bibinfo{title}{Performance limits on photonic heterostructures}}
  (\bibinfo{year}{2023}), \eprint{2307.00629}.

\end{thebibliography}


\begin{thebibliography}{2}
\expandafter\ifx\csname natexlab\endcsname\relax\def\natexlab#1{#1}\fi
\expandafter\ifx\csname bibnamefont\endcsname\relax
  \def\bibnamefont#1{#1}\fi
\expandafter\ifx\csname bibfnamefont\endcsname\relax
  \def\bibfnamefont#1{#1}\fi
\expandafter\ifx\csname citenamefont\endcsname\relax
  \def\citenamefont#1{#1}\fi
\expandafter\ifx\csname url\endcsname\relax
  \def\url#1{\texttt{#1}}\fi
\expandafter\ifx\csname urlprefix\endcsname\relax\def\urlprefix{URL }\fi
\providecommand{\bibinfo}[2]{#2}
\providecommand{\eprint}[2][]{\url{#2}}

\bibitem[{\citenamefont{Boyd}(2008)}]{boyd_nonlinear_2008}
\bibinfo{author}{\bibfnamefont{R.~W.} \bibnamefont{Boyd}},
  \emph{\bibinfo{title}{Nonlinear Optics}} (\bibinfo{publisher}{{Academic
  Press}}, \bibinfo{address}{{Amsterdam ; Boston}}, \bibinfo{year}{2008}),
  \bibinfo{edition}{3rd} ed., ISBN \bibinfo{isbn}{978-0-12-369470-6}.

\bibitem[{\citenamefont{Miller et~al.}(2016)\citenamefont{Miller, Polimeridis,
  Reid, Hsu, DeLacy, Joannopoulos, Solja{\v c}i{\'c}, and
  Johnson}}]{miller_fundamental_2016}
\bibinfo{author}{\bibfnamefont{O.~D.} \bibnamefont{Miller}},
  \bibinfo{author}{\bibfnamefont{A.~G.} \bibnamefont{Polimeridis}},
  \bibinfo{author}{\bibfnamefont{M.~T.~H.} \bibnamefont{Reid}},
  \bibinfo{author}{\bibfnamefont{C.~W.} \bibnamefont{Hsu}},
  \bibinfo{author}{\bibfnamefont{B.~G.} \bibnamefont{DeLacy}},
  \bibinfo{author}{\bibfnamefont{J.~D.} \bibnamefont{Joannopoulos}},
  \bibinfo{author}{\bibfnamefont{M.}~\bibnamefont{Solja{\v c}i{\'c}}},
  \bibnamefont{and} \bibinfo{author}{\bibfnamefont{S.~G.}
  \bibnamefont{Johnson}}, \bibinfo{journal}{Optics Express}
  \textbf{\bibinfo{volume}{24}}, \bibinfo{pages}{3329} (\bibinfo{year}{2016}),
  ISSN \bibinfo{issn}{1094-4087}.

\end{thebibliography}
\end{document}


\preprint{APS/123-QED}

\newcommand\numthis{\stepcounter{equation}\tag{\theequation}}
\newcommand{\im}[1]{\operatorname{Im}\left[#1\right]}
\newcommand{\re}[1]{\operatorname{Re}\left[#1\right]}

\title{Fundamental limits on $\chi^{(2)}$ second harmonic generation: supplemental document}
\date{July 2023}

\author{Jewel Mohajan}
\affiliation{Department of Electrical and Computer Engineering, Princeton University, Princeton, New Jersey 08544, USA}

\author{Pengning Chao}
\affiliation{Department of Electrical and Computer Engineering, Princeton University, Princeton, New Jersey 08544, USA}

\author{Weiliang Jin}
\affiliation{Flexcompute Inc., 130 Trapelo Road, Belmont, Massachusetts 02478, USA}

\author{Sean Molesky}
\affiliation{Department of Engineering Physics, Polytechnique Montréal, Montréal, Québec H3T 1J4, Canada}

\author{Alejandro W. Rodriguez}
\affiliation{Department of Electrical and Computer Engineering, Princeton University, Princeton, New Jersey 08544, USA}

\begin{abstract}
This document provides supplementary information to ``Fundamental limits on $\chi^{(2)}$ second harmonic generation". We sketch a derivation of the conservation of total power going into the cavity. We also show that enforcing passivity alone does not yield a useful bound, further illustrating the importance of constraints that account for wave scattering effects and cross-frequency constraints that connect fundamental and second harmonic fields to the same physical structure. 
\end{abstract}

\maketitle
\textbf{Total power conservation}---To derive the total ``resistive" power conservation across the fundamental and second harmonic frequencies, we consider Maxwell's wave equation at both the frequencies together~\cite{boyd_nonlinear_2008}:
\begin{align}\label{eq:MaxwellFFSH}
    \begin{bmatrix}
        \mathbb{M}_1 & 0 \\
        0 & \mathbb{M}_2
    \end{bmatrix}
    \begin{bmatrix}
        \vb{E}_1\\
        \vb{E}_2
    \end{bmatrix}
    =
    &\begin{bmatrix}
        \overline{\chi}_1 & \overline{\chi}^{(2)} \vb{E}_1^* \odot\\
        \overline{\chi}^{(2)}\vb{E}_1 \odot & \overline{\chi}_2
    \end{bmatrix}
    \begin{bmatrix}
        \vb{E}_1\\
        \vb{E}_2
    \end{bmatrix}
    +
    \begin{bmatrix}
        \frac{i}{\omega_1\varepsilon_0} \vb{J}^{inc}\\
        0
    \end{bmatrix}.
\end{align}
Assuming $\chi^{(2)}$ to be real and to have full permutation symmetry~\cite{boyd_nonlinear_2008} so that 
$\im{\vb{E}_1^\dagger \overline{\chi}^{(2)} \left(\vb{E}_1^* \odot \vb{E}_2 \right)} = -\im{\vb{E}_2^\dagger \overline{\chi}^{(2)} \left(\vb{E}_1 \odot \vb{E}_1 \right)}$,
we act on (\ref{eq:MaxwellFFSH}) with $\begin{bmatrix}
    \vb{E}_1^\dagger & \vb{E}_2^\dagger
\end{bmatrix}$ and taking the imaginary part of the equation to get:
\begin{equation}
\begin{aligned}
    &\im{\begin{bmatrix}
        \vb{E}_1^\dagger & \vb{E}_2^\dagger
    \end{bmatrix}
    \begin{bmatrix}
        \mathbb{M}_1 & 0 \\
        0 & \mathbb{M}_2
    \end{bmatrix}
    \begin{bmatrix}
        \vb{E}_1\\
        \vb{E}_2
    \end{bmatrix}}
    \\
    =
    &\begin{bmatrix}
        \vb{E}_1^\dagger & \vb{E}_2^\dagger
    \end{bmatrix}
    \begin{bmatrix}
        \overline{\chi}_1^a & 0\\
        0 & \overline{\chi}_2^a
    \end{bmatrix}
    \begin{bmatrix}
        \vb{E}_1\\
        \vb{E}_2
    \end{bmatrix}
    +
    \im{\vb{E}_1^\dagger \cdot \frac{i}{\omega_1 \varepsilon_0}\vb{J}^{inc}}
\end{aligned}
\end{equation}
where the superscript $(^a)$ denotes the anti-symmetric part of an operator, $\mathbb{O}^a = (\mathbb{O}-\mathbb{O}^\dagger)/2i$. Now, taking the down-conversion to be ``small" (technically, $\norm{\overline{\chi}^{(2)}(\vb{E}_1^* \odot \vb{E}_2)} \ll \norm{\overline{\chi}_1\vb{E}_1}$ under an appropriate definition of norm), we arrive at
\begin{equation}\label{eq:totalp_sup}
\begin{aligned}
    &\im{\int_\Omega \vb{E}_1^{{inc}*} \vdot \vb{P}_1 \ d\vb{r}} \\
    &= \int_\Omega \vb{P}_1^* \vdot \left( \mathbb{G}_1 + \chi_1^{-\dagger}\Id \right)^a \vb{P}_1 \ d\vb{r}
    + \int_\Omega \vb{P}_2^* \vdot \mathbb{G}_2^a \vb{P}_2 \ d\vb{r} \\
    &+ \int_\Omega \left(\vb{P}_2 - \vb{P}^{(2)}\right)^* \vdot (\chi_2^{-\dagger})^a \left( \vb{P}_2 - \vb{P}^{(2)} \right) \ d\vb{r},
\end{aligned}
\end{equation}
which states the fact that the ``resistive" power drawn from the input field is equal to the sum of all the radiative and absorptive powers across all the frequencies.

\textbf{Passivity bounds}--- The norm constraint on polarization stemming from passivity at the fundamental frequency relating to the incident field is given by~\cite{miller_fundamental_2016}:
\begin{equation}\label{eq:passivity1}
    \norm{\vb{P}_1} \leq \frac{\abs{\chi_1}^2}{\im{\chi_1}} \norm{\vb{E}_1^{inc}}
\end{equation}
where $\norm{\cdot}$ is the 2-norm for finite dimensional vector space and $\chi_1$ is the material susceptibility at $\omega_1$.

\begin{figure*}
  \includegraphics[width=\textwidth]{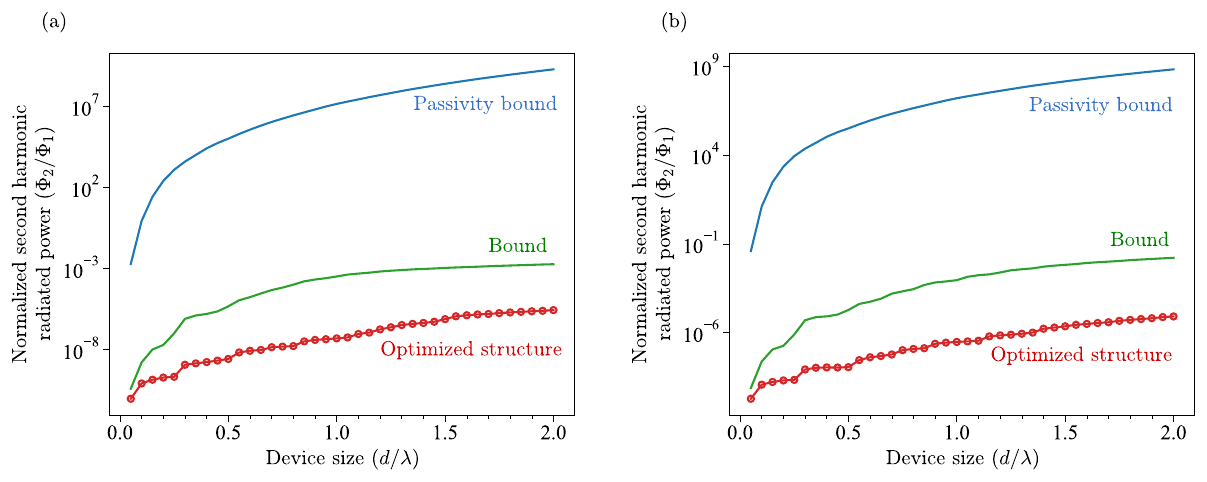}
  \caption{Comparison of bounds obtained from (\ref{eq:matlossbounds}) with bounds computed in the main text for (a) planewave incident normally, and (b) TM dipole source at the vicinity of the design region as source with the same material parameters and input powers used in the main text. Inverse designs (open circles) are also included for reference in both examples.} 
  \label{fig:passivity_bounds}
\end{figure*}

The norm constraint for non-linear up-conversion using~(\ref{eq:passivity1}), is given by
\begin{align}\label{eq:nl2}
    \norm{\vb{P}^{(2)}}^2 &= \abs{\frac{\chi^{(2)}}{\chi_{1}^2}}^2 \int_{\Omega} \vb{P}_1^{2*} \vdot \vb{P}_1^2 \ d\vb{r} \nonumber \\ &\leq \mathcal{R} 
    \abs{\frac{\chi^{(2)}}{\chi_{1}^2}}^2 \norm{\vb{P}_1}^4,
    \nonumber\\
    &\leq \mathcal{R} 
    \abs{\chi^{(2)}}^2 \frac{\abs{\chi_1}^4}{\left(\im{\chi_1}\right)^4} \norm{\vb{E}_1^{inc}}^4
\end{align}
where for finite difference representations the resolution factor is $\mathcal{R} = \frac{1}{\Delta v}$ with $\Delta v$ as the volume of an individual voxel.
Now, assuming the current from the up-conversion as a free source $\vb{J}_2 = -i\omega_2\varepsilon_0 \vb{P}^{(2)}$, the incident field at $\omega_2$ is given by $\vb{E}_2^{inc} = \frac{i}{\omega_2\varepsilon_0}\mathbb{G}_2 \vb{J}_2 = \mathbb{G}_2 \vb{P}^{(2)}$.
Finally, using passivity again at $\omega_2$ the linear polarization at the second harmonic is bounded by the relation $\norm{\vb{P}_2 - \vb{P}^{(2)}} \leq \frac{\abs{\chi_2}^2}{\im{\chi_2}} \norm{\vb{E}_2^{inc}}$, which in turn leads to:
\begin{equation}\label{eq:passivity2}
\begin{aligned}
    \norm{\vb{P}_2} &\leq \frac{\abs{\chi_2}^2}{\im{\chi_2}} \norm{\mathbb{G}_2} \norm{\vb{P}^{(2)}} 
    +
    \norm{\vb{P}^{(2)}}
    \\
    &\leq \mathcal{R}^{\frac{1}{2}} 
    \left( 1+ \frac{\abs{\chi_2}^2}{\im{\chi_2}} \norm{\mathbb{G}_2} \right)
    \frac{\abs{\chi^{(2)}}\abs{\chi_1}^2}{\left(\im{\chi_1}\right)^2} \norm{\vb{E}_1^{inc}}^2
    \\
\end{aligned}
\end{equation}
where we used the triangle inequality of norms ($\norm{x+y} \leq \norm{x}+\norm{y}$). Also note that, the norm $\norm{\mathbb{G}}$ is computed as the largest singular value of $\mathbb{G}$, i.e., square root of the largest eigenvalue of of $\mathbb{G}^\dagger \mathbb{G}$.
Therefore, the radiated power at the second harmonic is bounded by:
\begin{align}\label{eq:matlossbounds}
    \Phi_2&=\frac{1}{2}\omega_2 \varepsilon_0 \im{\int_\Omega \vb{P}_2^* \mathbb{G}_2 \vb{P}_2 d\vb{r}} 
    \nonumber\\
    &\leq \frac{1}{2}\omega_2 \varepsilon_0 \norm{\mathbb{G}^a_2} \norm{\vb{P}_2}^2
    \nonumber\\
    &\leq\frac{1}{2}\omega_2 \varepsilon_0 
    \norm{\mathbb{G}^a_2}
    \mathcal{R} \left( 1+ \frac{\abs{\chi_2}^2}{\im{\chi_2}} \norm{\mathbb{G}_2} \right)^2
    \frac{\abs{\chi^{(2)}}^2\abs{\chi_1}^4}{\left(\im{\chi_1}\right)^4} \norm{\vb{E}_1^{inc}}^4
\end{align}
In figure~\ref{fig:passivity_bounds}, we make a comparison of how far these passivity bounds are from the bounds calculated in the main text. It is clear that even after considering a high loss in the material ($\chi_1=\chi_2 = 3+0.1i$), the bounds calculated using just passivity are many orders of magnitude off from the bounds derived in the main text.

\bibliography{shg_refs}